# Laser microfluidics: fluid actuation by light

Jean-Pierre Delville[1], Matthieu Robert de Saint Vincent[1], Robert D. Schroll[2], Hamza Chraïbi[3], Bruno Issenmann[1], Régis Wunenburger[1], Didier Lasseux[3], Wendy W. Zhang[2] and Etienne Brasselet[1]

[1]Université Bordeaux 1, Centre de Physique Moléculaire Optique et Hertzienne, UMR CNRS 5798, 351 Cours de la Libération, F-33405 Talence cedex, France

[2]Physics Department and the James Franck Institute, University of Chicago, 929 East 57th Street, Chicago, Illinois 60637, USA

[3]Université Bordeaux I, Transferts, Écoulements, Fluides, Énergétique, UMR CNRS 8508, Esplanade des Arts et Métiers, F-33405 Talence Cedex, France.

E-mail: jp.delville@cpmoh.u-bordeaux1.fr

Abstract: The development of microfluidic devices is still hindered by the lack of robust fundamental building blocks that constitute any fluidic system. An attractive approach is optical actuation because light field interaction is contactless and dynamically reconfigurable, and solutions have been anticipated through the use of optical forces to manipulate microparticles in flows. Following the concept of "optical chip" advanced from the optical actuation of suspensions, we propose in this survey new routes to extend this concept to microfluidic two-phase flows. First, we investigate the destabilization of fluid interfaces by the optical radiation pressure and the formation of liquid jets. We analyze the droplet shedding from the jet tip and the continuous transport in laser-sustained liquid channels. In a second part, we investigate a dissipative light-flow interaction mechanism consisting in heating locally two immiscible fluids to produce thermocapillary stresses along their interface. This opto-capillary coupling is implemented in adequate microchannel geometries to manipulate two-phase flows and propose a contactless optical toolbox including valves, droplet sorters and switches, droplet dividers or droplet mergers. Finally, we discuss radiation pressure and opto-capillary effects in the context of the "optical-chip" where flows, channels and operating functions would all be performed optically on the same device.

## 1. Introduction

Fully integrated microfluidic laboratories [1], operating at nanoliter fluid volumes, are now recognized to have the potential for managing high-throughput massively multiplexed arrays of chemical microreactors or biological assays [2] by significantly reducing the material consumption, increasing data collecting rates, and allowing many different operations to be processed in serial or in parallel. Conversely, recent advances in manipulation of colloidal particles within microfluidic devices [3, 4] have shown that optical forces may represent a new and promising route toward microactuation. Indeed, beyond the fact that light fields can easily be focused to micrometric spots, they are contactless, dynamically reconfigurable, and do not impose any dedicated microfabrication. Contrary to classical total micro-analysis systems, which are generally constrained to perform pre-designed tasks, this optical flexibility proved to be very powerful to manipulate suspensions in a single fluid phase via optical forces, including pumping [5, 6, 7], sorting [8, 9, 10, 11], conveying [12]and propelling [13, 14]. Parallel actuation was also demonstrated [15]. Moreover, the combination of optics and microfluidics, called optofluidics [16, 17, 18], also became an emergent research field offering an unprecedented level of integration for building a new generation of optically-actuated microdevices merging optical reconfigurability, softness of fluid, smoothness of fluid interfaces, and compactness. Along this line, great achievements were realized in optical switching [19] and splitting [20], adaptive lensing [21, 22], interferometry [23], and lasing [24]. Then, using all these new ideas, we can address the following question: Are they routes to build "total-optical-lab-on-a-chip", where flows, channels and operating functions would all be performed optically on the same device? Glückstad [25] already advanced the concept of "optical chip", thinking of a chip where a suspension would be entirely actuated by light. The present survey aims at extending this concept to the optical manipulation of fluids and digital microfluidics, i.e. droplet microfluidics. It is indeed appealing to investigate how far light can (or cannot) be used to build channels, induce flows, drive droplet formation, and finally actuate droplets in channels.

Up to now, two types of devices emerged for the manipulation of drops at a microscale: open substrate and closed microchannels. In the first case, droplets are formed and manipulated on a surface or interface through the modulation of surface stresses by thermal [26], electrical [27], optical [12], photoresponsive [28] or chemical means [29]. Heat carried by a focused laser beam was also used to move droplets [30] or continuous fluid interfaces [31] on liquid or solid substrates, thus simplifying microfabrication since actuators are not embedded in the substrate anymore. This approach nevertheless suffers a major drawback for application purposes: since devices are open, reliability requires a well controlled environment to prevent cross contamination between the droplet and the atmosphere. Manipulation on solid substrates also leads to limitations in droplet velocity [32]. Conversely, in microchannels, drops are formed and transported in a closed chamber using a carrier fluid, usually oil for wetting purpose. In this case droplets are classically in the nanoliter range. A list of fundamental necessary operations was recently enumerated [33], including the need to block, merge, divide or sort droplets individually. Many solutions were proposed to divide or merge drops, to mix their contents and demonstrate their use as microreactors [34, 35]. However, due to the robustness of the environment of a microchannel, actuation functions are usually predetermined by the channel geometry and almost no further control is possible on individual droplets. Thus, a general and reliable solution for actively controlling droplets in microchannels still remains unavailable. The implementation of these functions would open the way to more complex operations being performed which may then be combined into a droplet based lab-on-a-chip. As illustrated above by the accurate manipulation of suspensions by lasers, an attractive route could be optical actuation.

The extension of particle manipulation to multiphase flows in bulk is nevertheless at a very early stage because bulk and surface optical forces are usually too weak to counteract hydrodynamic forces acting on fast moving droplets. The deformation of fluid interfaces by the optical radiation pressure is weak as well because interfacial tension generally behaves as a very efficient stabilizing effect. However, since droplet generation always involves fluid interfaces and since emulsification in microchannels always involves surfactant, it becomes necessary to re-investigate this coupling between liquid interface and laser waves. To do so and decrease the relative importance of interfacial tension versus optical radiation pressure, we considered a model system and performed experimental investigations on the meniscus of phase separated liquid mixtures close to a liquid-liquid critical point. Indeed, near-critical interfaces become highly deformable because interfacial tension vanishes close to a critical point. In these conditions, huge stationary interface deformations of several hundreds of microns can be generated with low beam power c.w. lasers. The induced hump can as well become unstable at sufficiently large beam intensities and lead to the formation of stationary beam-centered liquid micro-jet emitting droplets, anticipating the bases of our total-optical approach of microfluidics. However, as far as flowing microdroplets are used as tight fluid reactors or carriers, new fluid-handling micro-technologies are also required to convey, synchronize and more generally to manipulate accurately these droplets within a microchannel. To overcome optical radiation pressure limitations, we used instead the thermocapillary (or Marangoni) effect [36] induced by a laser. By locally heating the drop surface, an interfacial tension gradient is set up creating an imbalance in the surface stresses acting on the drop. In the low Reynolds-number regime of microfluidics, this affects the bulk flow very rapidly and allows for the implementation of an optical toolbox for total control of digital microfluidics.

The paper is organized in three main parts. After a small survey on optical manipulation of liquid interfaces, we show in Section 2 how a laser wave can couple to a liquid interface and drive its bending and how this bending can become unstable and trigger a liquid jet emitting microdroplets continuously. We investigate this liquid flow and demonstrate that it originates from bulk radiation pressure effects. Finally, we show how this "opto-hydrodynamic" instability can be used to form and stabilize large aspect ratio liquid channels to carry liquids. In Section 3, we switch to the optical actuation of droplets by thermocapillary forces in microchannels and demonstrate its potential to create various drop-handling functions. In the fourth prospective Section, an attempt at combining radiation pressure and thermocapillary effects is discussed in order to dream about the "total-optical-lab-on-a-chip" concept for further appealing investigations of contactless digital microfluidics.

## 2. Optical manipulation of fluid interfaces and light-induced jetting

### 2.1. Mechanical aspects

The first experiment on deformation of liquid interface by the radiation pressure of a laser wave [37] is contemporary to the optical levitation demonstration (they were both performed by the same authors). Considering some earlier theoretical developments [38], the first motivation of Ashkin and Dziedzic was to determine in which direction occurs the bending of a fluid interface separating two dielectric liquids of different refractive indices. Their result was sufficiently non intuitive to briefly summarize them.

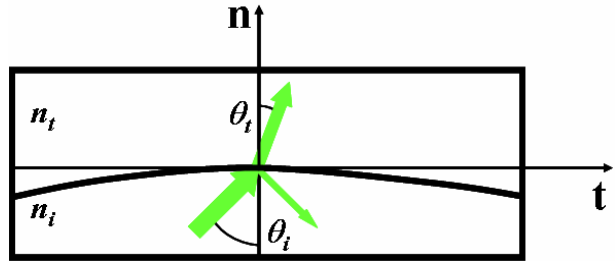

Figure 1: Schematic representation used for the calculation of the radiation pressure exerted by a laser beam on an interface separating two dielectric fluids characterized by the indices of refraction $n_i$ and $n_t$. The angle of incidence and transmission are respectively $\theta_i$ and $\theta_t$. **t** and **n** are the tangent and the normal directions at the location where the beam impinges the interface.

Let us consider two dielectric media, of different refractive indices $n_{i,t}$, separated by an interface (Figure 1). The subscripts $i$ and $t$ refer to incidence and transmission and $\theta_i$ and $\theta_t$ are respectively the incident and the transmitted angles. Since the momentum carried by the incident propagating light depends on the refractive index, it is not conserved when the beam travels from one to the other dielectric and the resulting discontinuity gives birth to a radiation pressure applied to the interface. Considering the Minkowski formalism [39], its expression is given by:

$$\mathbf{\Pi}_{Rad} = n_i \cos^2 \theta_i \left[ 1 + R(\theta_i, \theta_t) - \frac{\tan \theta_i}{\tan \theta_t} T(\theta_i, \theta_t) \right] \frac{I}{c} \mathbf{n}, \qquad (1)$$

where $I$ is the beam intensity and, $R(\theta_i, \theta_t)$ and $T(\theta_i, \theta_t) = 1 - R(\theta_i, \theta_t)$ are the classical reflection and transmission Fresnel coefficients in electromagnetic energy. Consequently, $\mathbf{\Pi}_{Rad}$ is always normal to the interface. Moreover, by considering the expressions of the reflection and transmission coefficients $R(\theta_i, \theta_t)$ and $T(\theta_i, \theta_t)$ [40], it can be easily shown that the optical radiation force $\mathbf{\Pi}_{Rad}$ is always directed toward the dielectric medium of lowest index of refraction. This means that interface bending does not depend on beam propagation, and the main reason for that is that photon gains momentum when passing from a low to a large refractive index medium. Ashkin and Dziedzic [37] verified this remarkable property which was further generalized to the liquid-liquid case [41]. A quantitative interpretation was nevertheless difficult due to the spatial and temporal profiles of the laser pulses used [42]. For instance, Brevik [43] had to consider a larger beam waist and to use a time delay to retrieve the temporal behavior of the induced lens. In fact, very few results related to the deformation of fluid interfaces by the optical radiation forces have been published. Except a brief letter [44], we are aware of a second set of experiments realized at the liquid/air interface [45] using a ruby laser. As interface bending was still very weak, detection was performed holographically.

Nevertheless, the emergence of the so-called soft matter physics in the eighties brought a sort of renewal of laser radiation pressure by opening new horizons, particularly in metrology of soft interfaces. For instance, a Japanese team exploited the bending of an interface by a laser wave to measure interfacial tensions [46] in a non-contact manner. In a second work [47], they focused their attention on the difficult case of ultra low interfacial tension, for which most of classical techniques fail. Finally, they showed the pertinence of radiation pressure to measure shear viscosities [48]. The characterization of the mechanical properties of lipidic membranes is also of fundamental interest in biophysics. Since optical dynamometry is limited to large vesicles, Wang's group used radiation pressure to quantitatively measure curvature rigidity of small vesicles [49] and in vivo cells [50] at a nanometric length scale. Compared to optical dynamometry using a beam-trapped bead as a probe [51], these experiments show that the main interests in using radiation pressure are the strong localization of excitation as well as its contactless character. In the meantime, Käs' group patented a new tool, called "Optical Stretcher", to probe the elasticity of red blood cells using the radiation pressure [52]. Due to its sensitivity, one of the ultimate goals of this technique is to differentiate healthy from malignancy cells from the difference in elastic response [53, 54].

Conversely, Zhang and Chang [55] performed the first direct observations of laser-induced nonlinear interface deformation on spherical water droplets using a high speed camera. After the travel of a *100 mJ* pulse, a first series of images evidenced that interface deformation is directed toward the air on both sides of the drop with a small asymmetry on the exit face as a result of the intensity increase due to the ball lens behavior of the drop. A quantitative interpretation of this weakly nonlinear regime of deformation was published by Poon and coworkers [56] and further developed by Brevik and Kluge [57].Even more surprising, Zhang and Chang showed that *200 mJ* pulses lead to droplet disruption and the formation of a long liquid filament on the exit face. Note that nonlinear interface deformations were also observed on vesicles using c. w. lasers [53].

### 2.2. Principles of the experiment

As discussed above, deformations induced by a continuous laser beam on classical interfaces remain weak. Thus, direct visualization of deformations induced by a c.w. laser obviously requires very soft interfaces. We thus choose to work with near-critical interfaces because the interfacial tension $\sigma$ vanishes close to a critical point as $\sigma = \sigma_0 \left| T/T_C - 1 \right|^{1.26}$, where $T$ and $T_C$ are respectively the actual and critical temperatures. Moreover, given a temperature shift to criticality, the use of supramolecular liquids leads to even smaller interfacial tension. This is due to the fact that the amplitude $\sigma_0$ is inversely proportional to the square of the characteristic molecular length scale $\xi_0$ according to a universal ratio $R^+$, defined as

$R^+ = \left( \sigma_0 \xi_0^2 / k_B T_C \right) = 0.39$ [58] resulting from the renormalization group theory. Consequently, in order to enhance radiation pressure effects on fluid interfaces, we performed experiments in near critical phase separated micellar phase of microemulsions (assembly of surfactant-coated water nano-droplets suspended in oil) [41]. As they belong to the same universality class ($d = 3$, $n = 1$) of the Ising model like all isotropic liquids, results obtained from our "exotic" choice of fluid can be transposed to any interface separating liquids belonging to the same universality class. We used a quaternary mixture of sodium dodecyl sulfate (surfactant), water, toluene (oil) and n-butanol-1 (alcohol). Its mass composition in weight fraction is *4 %* SDS, *9 %* water, *70 %* toluene, and *17 %* n-butanol-1. This composition was chosen so as to be critical at a temperature $T_C = 35\ °C$ [59]. As illustrated in the Inset of Figure 2, for a temperature $T > T_C$, the mixture phase separates in two micellar phases of different concentration $\Phi$, called hereafter $\Phi_1$ and $\Phi_2$. For this composition, the micelle radius is $\xi_0 = 40 \pm 2\ \mathring{A}$, i.e. one order of magnitude larger than classical molecular length scale, and small enough to get a suspension transparent in the visible optical wavelength window. Due to the amplitude of this $\xi_0$ value, the interfacial tension is expected to be extremely weak compared to that of liquid-vapor interfaces. We found $\sigma$ to be one million times smaller than that of the water free surface ($\sigma = 72\ mN/m$) at *1.5 K* from the critical point. This weakness clearly justifies the reason why we expect large interface deformations using c.w. lasers. As discussed below, other quantities either vanish (density and refractive index contrast) or diverge (correlation length of density fluctuations and osmotic compressibility) close to the critical point.

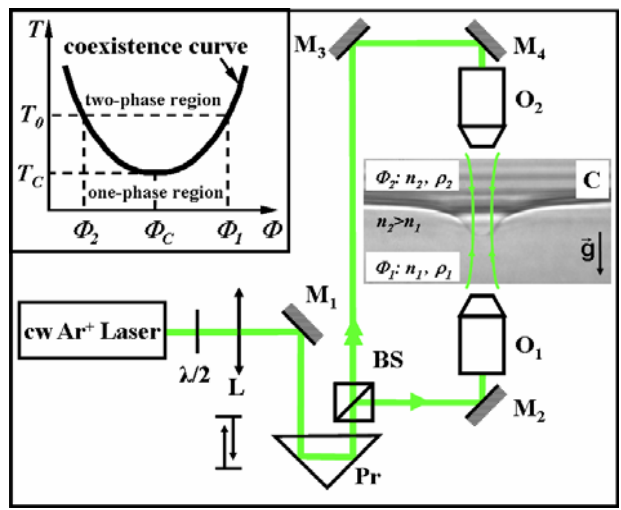

Figure 2 (adapted from Ref. [60]) : Experimental setup. $\lambda/2$ : half-wave plate, $L$ : lens, $M_{i=1,4}$ : mirrors, $Pr$ : prism; $BS$ : beam splitter; $O_{i=1,2}$ microscope objectives *10×*. $C$ denotes the experimental cell in which an example of deformed near-critical interface separating the phases $\Phi_1$ and $\Phi_2$ for $T_0 > T_C$ is presented. Inset: schematic phase diagram of the used micellar phase of microemulsion. $T$ is temperature and $\Phi$ is the volume fraction of micelles.

Finally, the optical absorption of our micellar phases at the used wavelength is $\alpha_a = 3.10^{-4}\ cm^{-1}$. Thus thermal heating and associated side effects do not disturb interface deformation and will be discarded.

The experimental setup is presented in Figure 2. As beam intensity depends on both power and beam radius, the main point is to locate the beam waist $\omega_0$ on the meniscus of the phase-separated mixture and to vary its size. This was done by using the lens $L$ to form a first intermediate waist. The variation of the beam waist $\omega_0$ is performed by moving the prism $Pr$ to increase the optical path between $L$ and the focusing long working distance microscope objectives, either $O_1$ or $O_2$ ($10\times$). As the corresponding beam waist position varies within the cell, $C$ is mounted on translation stages. Moreover, our setup allows for ascending and descending beams by using the half-wave plate $\lambda/2$ and the beam splitter $BS$. Interface deformations are induced by the radiation pressure of a continuous $TEM_{00}$ Argon-Ion laser (Innova 90, Coherent, working wavelength $\lambda_0 = 0.514\ \mu m$). Using this procedure accessible values are $\omega_0 = 4.8 - 32.1\ \mu m$ (resp. $\omega_0 = 3.5 - 11.3\ \mu m$) for ascending (resp. descending) beams. Considering weakly focused beams, we assume the beam divergence to be negligible around the focus and write the intensity profile as:

$$I\left(r,z\right) \approx I\left(r\right) = \frac{2P}{\pi\omega_0^2} exp\left(-\frac{2r^2}{\omega_0^2}\right), \tag{2}$$

where $P$ is the incident beam power. Strictly speaking, this approximation is correct as far as $\lambda_0 h / \left(n\pi\omega_0^2\right) < 1$, where $h$ and $n$ are respectively the height of the deformation and the index of refraction. Observations of deformations are performed transversally using a white light source for illumination and a microscope for imaging on a C.C.D. video camera coupled to a frame grabber. A spectral filter is placed between the microscope and the C.C.D. to eliminate residual laser light scattered by the micellar phases.

### 2.3. Light-induced jetting

The general expression governing the stationary interface shape $h\left(r\right)$ under radiation pressure effects is given in cylindrical coordinates by [41]:

$$\left(\rho_1 - \rho_2\right)gh\left(r\right) - \sigma\kappa\left(r\right) = \Pi_{Rad}\left(r\right), \tag{3}$$

where $\kappa(r)$ is the interface double mean curvature. Equation (3) shows that the height of the deformation results from the balance between radiation pressure $\Pi_{Rad}(r)$, given by Eq. (1), and both buoyancy $(\rho_1 - \rho_2)gh(r)$ and Laplace pressure $\Pi_{Laplace}(r) = -\sigma\kappa(r)$. It leads:

$$(\rho_1 - \rho_2)gh(r) - \frac{\sigma}{r}\frac{d}{dr}\left(\frac{rh'(r)}{\sqrt{1+h'(r)^2}}\right) = \frac{n_1 I(r)}{c}\cos^2\theta_i\left(1 + R - \frac{tan\theta_i}{tan\theta_t}T\right). \qquad (4)$$

The incidence and transmission angles $\theta_i$ and $\theta_t$ can be related to the shape of the deformation. For a downward beam, i.e. propagating from the large to the low refractive index medium, they are given by:

$$\begin{aligned}
cos\theta_i &= \frac{1}{\sqrt{1+h'(r)^2}}, \\
cos\theta_t &= \sqrt{1 - \left(\frac{n_2}{n_1}\right)^2 \frac{h'(r)^2}{1+h'(r)^2}}.
\end{aligned} \qquad (5)$$

Note that one can use the expression of $R$ and $T$ at normal incidence for weak deformations, and linearize curvature. In these conditions Equation 4 reduces to

$$(\rho_1 - \rho_2)gh(r) - \sigma\Delta_r h(r) = \frac{2n_2}{c}\left(\frac{n_1 - n_2}{n_1 + n_2}\right)I(r). \qquad (6)$$

While for an upward beam we observe a continuous transition from a bell to a nipple-like shape of increasing pedestal [61], the morphology is totally different when the beam propagates downwards, from the large to the low refractive index fluids. As illustrated in Figure 3, the deformation height deviates from the linear regime (shown by the dashed line) for increasing incident beam powers and diverges at some well-defined beam power threshold $P_\uparrow$. Above this instability threshold a beam centered stationary liquid jet forms, which emits droplets at its tip.

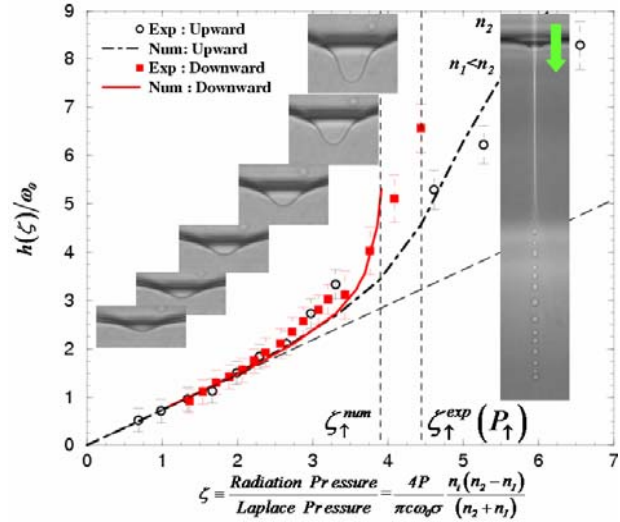

Figure 3 (adapted from Ref. [62]): Experimental and numerical evolution of the deformation height $h(r=0)/\omega_0$ versus the rescaled downward and upward radiation pressure $\zeta$ for $(T-T_C)=3\ K$ and $\omega_0=5.3\ \mu m$. The dashed line indicates the linear regime in deformation and the measured threshold $P_\uparrow$, above which the interface becomes unstable, is indicated by two vertical dashed lines, the experimental and the predicted one. Snapshots show the downward evolution for beam power increasing from left to right as $P=190$, $250$, $280$, $310$, $340$, $370$, and $P_\uparrow=400\ mW$ for the stationary jet. The total height of right picture is $1\ mm$.

In fact, interface instability can only be triggered by light if the incident fluid has the largest index of refraction. Indeed, in this case, total internal reflection of light can occur within the deformed interface which in turn increases light intensity and radiation pressure at the tip of the deformation, and so on [63]. Once the jet is formed, droplets are continuously shed at the tip. Since the index of refraction of the droplet is larger than that of the surrounding fluid, the beam automatically traps them. This brings directionality in emission and transport of droplets that can be actuated by tilting the beam. Moreover, beyond the continuous emission of liquid droplets, the mean jet length can be continuously tuned by increasing beam power [60]. Finally, at fixed $(T-T_C)$ and $\omega_0$, the droplet flow rate and the drop emission frequency can be controlled by the incident beam power as illustrated respectively in Figure 4 and 5 for $(T-T_C)=12\ K$ and $\omega_0=4.5\ \mu m$. Flow rate of hundreds of $\mu m^3/s$ are easily acheived with a shedding droplet frequency of several $Hz$. Consequently, contrary to other methods (flow mediated: [64, 65] or electrically mediated [66]), this optical approach of microfluidics provides droplets that are directly produced at a chosen micrometric size in a contactless way without further processing. Also, as droplets are produced in three dimensions from a jet size controlled by the beam, neither particular microfluidic device nor dedicated equipment is required to manage the droplet flow.

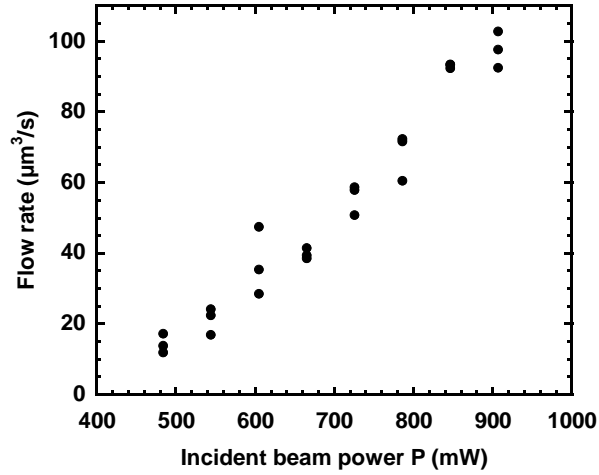

Figure 4: Flow rate variation within the jet versus beam power $P$ for $(T - T_C) = 12\ K$ and $\omega_0 = 4.5\ \mu m$.

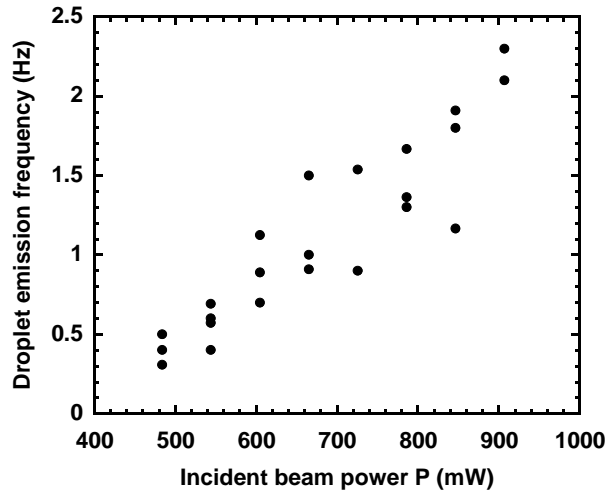

Figure 5: Droplet emission frequency variation at the jet tip versus beam power $P$ for $(T - T_C) = 12\ K$ and $\omega_0 = 4.5\ \mu m$.

### 2.4. Light-induced liquid channel

Classically, liquid columns of large aspect ratio are not stable. Indeed, beyond a certain aspect ratio, they are known to break into droplets due to the Rayleigh-Plateau instability [67]. Noticeable is the fact that laser waves are able to stabilize such structures that would be unstable otherwise. This can be easily demonstrated by removing the optical excitation; Rayleigh-Plateau rupture of stationary jets occurs instantaneously. This optical stabilization of jets advances the possibility to stabilize real liquid columns under intense illumination. We can indeed form liquid columns between the liquid interface and the bottom of the experimental cell. Thus, as the experimental optical cell is one or two millimeters high and the meniscus is located in the middle due to near criticality, it should be possible to connect the liquid jet to the bottom of the cell and to form a liquid column of super large aspect ratio. This is illustrated in Figure 6 in a one-millimeter thick cell. The corresponding aspect ratio is $\Lambda = L/2R \approx 70$, well above the Rayleigh-Plateau onset value $\Lambda = \pi$; $\Lambda$ values largely above $100$ are easily achieved in $2\ mm$ cells. Finally, as already demonstrated by the droplets shedding at the jet tip, hydrodynamic flow still persists within liquid channels as illustrated by the fluid accumulation at their feet.

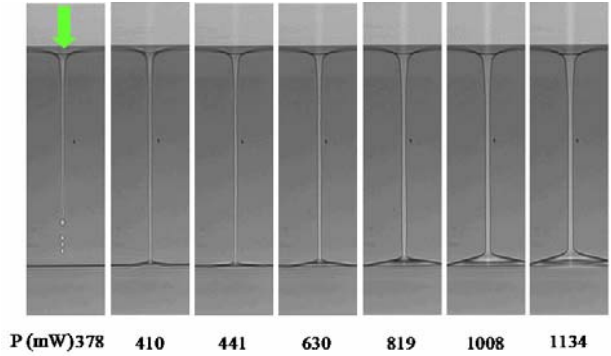

Figure 6: Formation of a liquid channel and increase of its diameter with the beam power. Parameters are $\omega_0 = 3.47~\mu m$ and $\left(T - T_C\right) = 4~K$. The liquid channel length is $334~\mu m$.

Since the optical specificity of liquid channels is to intrinsically behave as waveguides, as illustrated in Figure 7a, we suspected this guiding to be at the origin of the observed optical stabilization well above the Rayleigh-Plateau threshold. To check this hypothesis, we used in a first step a ray optics description of the light trapped in the bridge to investigate the balance between the radiation pressure of the guided photons and the Laplace pressure [68]. The model consists in counting the number of photons trapped within the channel by total reflection and to calculate the corresponding radiation pressure. Due to the funnel shape of the entrance of the channel, we assumed that the total power trapped in the channel, denoted $P_A$, equals the power shined on the interface between the channel radius $R$ and $\alpha R$ ($\alpha > 1$). As illustrated in Figure 7b, the parameter $\alpha$ is defined such that rays impinging on the funnel at radius $\alpha R$ have an impact angle of $\theta_{TIR}$, the total internal reflection angle; in a geometrical optics approach, the light shining directly into the bridge, i.e. at radius smaller than $R$, will not interact with the walls.

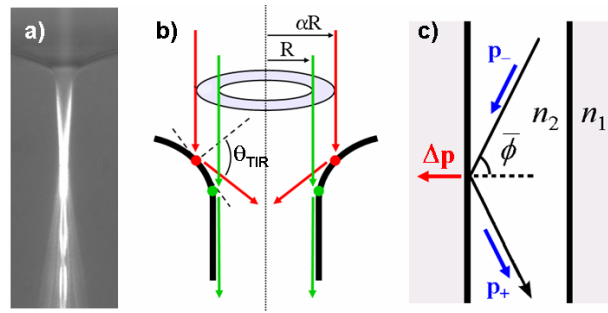

Figure 7 (adapted from Ref. [68]): Geometrical model for optical stabilization of liquid channels. (a) Light guiding by the channel observed by scattering; $P = 460~mW$, $\omega_0 = 3.47~\mu m$ and $\left(T - T_C\right) = 5~K$. (b) Rays participating to the stabilization of a liquid channel. (c) Linear momentum transfer $\Delta p$ of a photon that undergoes total internal reflection inside the bridge.

In experiments, we found $\alpha_{exp} = 2.13$ for $\omega_0 = 6.95 \ \mu m$, a value which remains almost constant within the investigated range of waists. Following this scheme, we find:

$$P_A = P \ exp\left(-\frac{2R^2}{\omega_0^2}\right) \times \left[1 - exp\left(-\frac{2\left(\alpha^2 - 1\right)R^2}{\omega_0^2}\right)\right].$$ (7)

Once the reflected light enters the bridge, it will bounce down the channel at some angle $\theta > \theta_{TIR}$ to the normal of the interface. Since the optical indices of the two fluids are close together in the vicinity of the critical point, $\theta_{TIR}$ almost equals $\pi/2$. More precisely, at $\left(T - T_C\right) = 4 \ K$, we have $\theta_{TIR} = 81.4°$. Thus, a reasonable approximation consists in assuming that all the guided light reflects at the same mean angle $\overline{\phi}$ and, further, that this angle can be approximated as $\overline{\phi} = \frac{1}{2}\left(\theta_{TIR} + \frac{\pi}{2}\right)$. As each guided photon of energy $E_\gamma$ reflects off of the channel interface, its linear momentum changes by $\Delta p = \left\|\mathbf{p}^+ - \mathbf{p}^-\right\| = 2\frac{n_2}{c}E_\gamma \ cos \ \overline{\phi}$. The channel interface consequently receives an equal impulse in the outward normal direction (see Figure 7c). Then, the total radiation pressure on the channel surface is the product of the momentum change per photon, $\Delta p$, by the number of reflections per unit area of the channel wall. This reflection density is given by the rate of photons $P_A/E_\gamma$ divided by the area per reflection for a single photon, $4\pi R^2 \ tan \ \overline{\phi}$. The resulting radiation pressure is:

$$\Pi_{Rad} = \frac{P_A/E_\gamma}{4\pi R^2 \ tan \ \overline{\phi}} \Delta p = \frac{n_2}{2c} \frac{cos \ \overline{\phi}}{tan \ \overline{\phi}} \frac{P_A}{\pi R^2}.$$ (8)

We can now deduce the bridge radius as a function of the laser power by balancing the radiation pressure $\Pi_{Rad}$, which tends to expand a cylindrical channel outwards, against the inward constricting stresses exerted by the Laplace pressure $\Pi_{Laplace} = \sigma/R$:

$$\frac{\sigma}{R} = \frac{n_2 - n_1}{4\pi c} \frac{P_A}{R^2}.$$ (9)

Figure 8 shows the predictions of the model compared to experimental data for $\left(T - T_C\right) = 4 \ K$ and $\omega_0 = 6.95 \ \mu m$. For an easy comparison, we have rescaled the channel diameter $d = 2R$ and the beam power $P$, respectively by the beam waist $\omega_0$ and the laser power $P_C$ below which no bridge can exist. This simple geometric model successfully captures all the qualitative features of laser-stabilized liquid channels formation. In particular, it demonstrates that a liquid column can be stabilized by light above a beam power threshold and retrieves the gradual widening of the channel radius with $P$. However, even if right orders of magnitude are found, there is, as illustrated in Figure 8, a quantitative mismatch in both the size of the bridge and the value of $P_C$. These variations can be explained by the use of a ray optics approach and its subsequent ray selection mechanism of the photons at the entrance of the channel which represents a partial picture of a more complicated propagation problem.

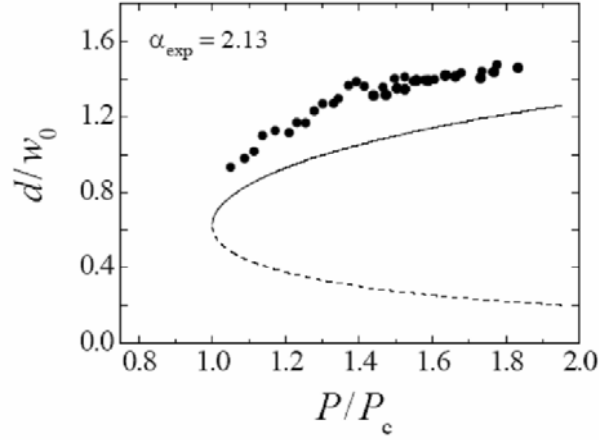

Figure 8 (adapted from Ref. [68]): Comparison between the prediction from a geometric model and experimental results obtained for $(T - T_C) = 4\ K$ and $\omega_0 = 6.95\ \mu m$. Solid (resp. dashed) line corresponds to stable (resp. unstable) states and circles are experimental results.

The mismatch as well as refined predictions were solved using a fully developed electromagnetic description [69, 70]. To do so, we consider the liquid channel as a semi-infinite cylinder of radius $R = d/2$ perfectly aligned with the incident beam, the inner and outer indices of refraction being respectively $n_2$ and $n_1$. Due to the use of near-critical fluids, $n_2$ and $n_1$ are very close together, and the modes propagating within the channel can be satisfactorily described by the linearly polarized $LP_{lm}$ modes [71]. Moreover, since $n_2 \simeq n_1$, only the first-order terms in $n_2 - n_1$ are retained in calculations. Since the incident beam is Gaussian, there is no coupling with $l > 1$ modes and only the $LP_{0m}$ modes are excited, each of them carrying a power $P_m = T_M P$, where $T_M$ is the coupling coefficient. The transmission for the $m$th mode is calculated from the normalized overlap integral between the incident Gaussian field and the propagating $LP_{0m}$ field:

$$T_m = \frac{8}{\omega_0^2 R^2} \frac{\left| \int_0^\infty \Re_m(r) \exp\left(-r^2/\omega_0^2\right) r dr \right|^2}{\dfrac{J_1^2(\kappa_m R)}{J_0^2(\kappa_m R)} + \dfrac{K_1^2(\gamma_m R)}{K_0^2(\gamma_m R)}}, \tag{10}$$

where $J_m$ and $K_m$ are the usual $m$th order Bessel functions and $\kappa_m$ and $\gamma_m$ are the $m$th roots of the characteristic equation defining the $LP_{0m}$ mode, $\kappa_m \dfrac{J_1(\kappa_m R)}{J_0(\kappa_m R)} = \gamma_m \dfrac{K_1(\gamma_m R)}{K_0(\gamma_m R)}$, with $\kappa_m{}^2 + \gamma_m{}^2 = \left(2\pi/\lambda_0\right)^2 \left(n_2{}^2 - n_1{}^2\right)$ [71]. The function $\Re_m(r)$ is associated to the radial profile of the electric field of the $m$th mode:

$$\Re_m(r) = \left( \left. \frac{J_0(\kappa_m r)}{J_0(\kappa_m R)}\right|_{r \leq R}, \ \left. \frac{K_0(\gamma_m r)}{K_0(\gamma_m R)}\right|_{r \geq R} \right). \tag{11}$$

For steady and perfectly cylindrical liquid channels in weightless conditions, the interface equilibrium condition writes, as before, $\Pi_{Rad} = \Pi_{Laplace}$, since the viscous stress due to a possible flow within a perfect cylinder (see subsection 2.5) is tangential to the interface. By definition, the radiation pressure $\Pi_{Rad}$ corresponds to the radial discontinuity of the electromagnetic stress tensor [72] across the normal to the interface. In the present situation involving nonmagnetic fluids, $\Pi_{Rad}$ writes at the lowest order in $n_2 - n_1$:

$$\Pi_{Rad} = \frac{1}{2}\varepsilon_0 \overline{n}\left(n_2 - n_1\right)\sum_m \left|\mathbf{E}^{(\mathbf{m})}\right|^2_{r=R}, \tag{12}$$

where $\overline{n} = \left(n_1 + n_2\right)/2$ is the averaged refractive index, $\varepsilon_0$ the permittivity in vacuum, and $\mathbf{E}^{(\mathbf{m})}$ the complex electric field of the $m$th mode. Then, since the flux of the Poynting vector across a plane perpendicular to the propagation axis $z$ equals the power carried by the $m$th mode of power $P_m = T_M P$, we deduce:

$$\left|\mathbf{E}^{(\mathbf{m})}\right|^2_{r=R} = \frac{2P_m}{\varepsilon_0 \pi c \overline{n} R^2}\left[ \frac{J_1{}^2(\kappa_m R)}{J_0{}^2(\kappa_m R)} + \frac{K_1{}^2(\gamma_m R)}{K_0{}^2(\gamma_m R)}\right]^{-1}. \tag{13}$$

Finally $\Pi_{Rad}$ is explicitly obtained by combining Equations 10-13 and the liquid channel diameter is further found by numerically solving the equilibrium equation $\Pi_{Rad} = \Pi_{Laplace}$. As illustrated in Figure 9, the model predicts that there is no equilibrium radius below a critical power and the emergence of the set of discrete solutions $\left\{R_{eq}^{(n)}\right\}$. The stability criterion is obtained from a standard static stability analysis at $R = R_{eq}^{(n)}$ and leads to $\partial \Pi_{Rad}/\partial R < \partial \Pi_{Laplace}/\partial R$. Further details in the calculation procedure can be found in Ref. [70].

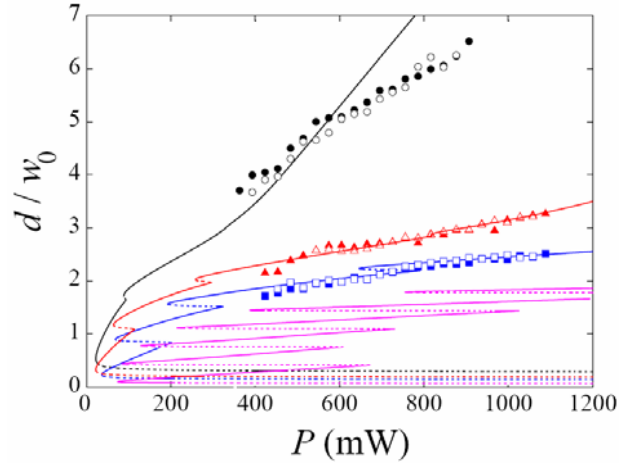

Figure 9 (adapted from Ref. [69]): Power dependence of the rescaled liquid channel diameter for different beam waist values: $\omega_0 = 1.8\ \mu m$ (black line and circles), $\omega_0 = 2.7\ \mu m$ (red line and triangles), $\omega_0 = 3.5\ \mu m$ (blue line and squares), and $\omega_0 = 7\ \mu m$ (magenta). Solid and open symbols refer to experimental data for increasing and decreasing beam power variations, respectively. Solid and dashed lines correspond respectively to stable and unstable states predicted by the model.

Results are displayed in Figure 9, where the solid (resp. dashed) line refers to stable (resp. unstable) solutions. For smallest waists, a good quantitative agreement with experimental data is obtained, particularly if we recall that there is no adjustable parameter. A more complex sequence of stable and unstable states is predicted at low beam powers, but this region cannot be explored experimentally because the large aspect ratio column detaches from the wetting film at the bottom of the cell. Note finally that a single and well-defined diameter does not imply a monomodal behavior. For example, the numbers of modes involved in the range of parameters presented in Figure 9 are $m_{max} = 6$ for $\omega_0 = 1.8\ \mu m$ and $m_{max} = 4$ for $\omega_0 = 2.7\ \mu m$. Conversely, when the beam waist is larger, the model predicts instead multistable states, as illustrated in Figure 9 for $\omega_0 = 7\ \mu m$. There are indeed several stable radii at fixed power that should be characterized experimentally. Then, one would expect to observe metastable states as well as a hysteretic behavior of $d(P)$. We rather found a diameter which varies along the column and with time. In fact, the column explores different discrete diameter values at a single time, which is also a signature of a bistable (or multistable) process [69]. Since the mean amplitude of thermally activated interface fluctuations $\left(k_B T / \sigma\right)^{1/2}$ [73], where $k_B T$ is the thermal energy, are within the range $0.1 - 0.2\ \mu m$, i.e. one order of magnitude smaller that the interplateaux distance, we eliminate their direct contribution to the observed spatial and temporal radius variations. However, the coupling between these thermal fluctuations with the light-induced flow inside the liquid column discussed in the following subsection 2.5, can also give birth to a normal viscous stress allowing exploration of multivalued fiber radii at large beam diameters. In conclusion, Figure 9 shows that the mechanism by which liquid channels can be sustained by light is clearly identified.

*2.5. Light-induced bulk flow*

We showed above how liquid jets can be driven and stabilized by light pressure. However, this surface force always applies normally to interfaces and thus, is unable to explain the fluid flow inside the structure at the origin of the droplet shedding at the jet tip or the fluid accumulation at the bottom of liquid channels. A bulk effect is required in the absence of tangential forcing. Radiation pressure is nonetheless known to push suspended particles, as demonstrated by optical levitation [74], because light momentum changes direction when it is elastically scattered by the particles. This momentum change results in a scattering force applied to the particle. As a macroscopic consequence, the photon momentum lost during the beam propagation in a suspension of nanoparticles is transferred to the fluid by momentum conservation. Light is thus able to drive bulk flows in liquid suspensions and emulsions. In a critical fluid, the correlation length of density fluctuations $\xi^+$, which diverges close to the critical point, plays this role of refractive index non-homogeneities [75], as schematically illustrated in the left Inset of Figure 10. The length scale of these density fluctuations being smaller than *100 nm* in experiments, we assumed them to behave as Rayleigh scatterers and to undergo individually a scattering force in the same manner as suspended particles. If we neglect the resulting intensity attenuation due to the associated fluid turbidity, the scattering force density is:

$$\mathbf{F_{Scat}}(r) = \frac{n_l I(r)}{\upsilon c} \oint_{\Omega} \left( \hat{\mathbf{k}}_0 - \hat{\mathbf{k}}_{Scat} \right) \Sigma \left( \mathbf{k}_{Scat} \right) d\Omega = \frac{\pi^3}{\lambda_0^4} \frac{n_l I(r)}{c} \left( \Phi \frac{\partial \varepsilon}{\partial \Phi} \right)^2 k_B T \chi_T f(\beta) \mathbf{z}, \qquad (14)$$

where $\hat{\mathbf{k}}_0$ and $\hat{\mathbf{k}}_{Scat}$ are the directions of incident and the scattered wave vector, $\upsilon^{-1}$ the number of scatterers per unit volume, $\varepsilon = n^2$, $\Sigma \left( \mathbf{k}_{Scat} \right)$ the scattering cross section, $\Omega$ the solid angle, $\chi_T$ the osmotic compressibility which also diverges close to the critical point, and $f(\beta)$ a function of $\beta = 2 \left( 2\pi n \xi^+ / \lambda_0 \right)^2$ [76]. Assimilating a jet to a fluid cylinder, and assuming that the low Reynolds number flow within this jet is driven by the scattering force, the Poiseuille equation leads to the following flow rate $Q = \frac{\pi}{128\eta} \left| \mathbf{F_{Scat}} \right| d^4$ for flows bounded by solid cylindrical boundaries. Despite this last approximation, the right order of magnitude is retrieved compared to the measured droplet shedding [75], illustrated in Figure 4.

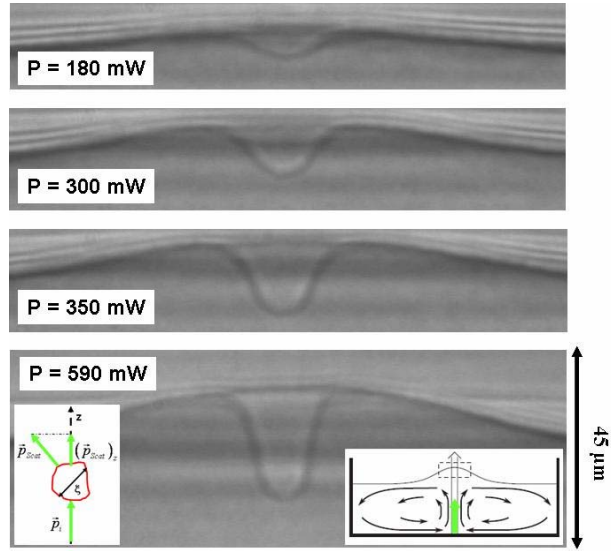

Figure 10 (adapted from Ref. [75]): Interface deformations when the laser shines upwards for $\left(T - T_C\right) = 1.5\ K$ and $\omega_0 = 8.9\ \mu m$. The interface forms a downward deformation (due to radiation pressure) and an upward, broad hump along beam axis. The left Inset illustrates the scattering mechanism for momentum transfer to the fluid and the right Inset presents the toroidal recirculation produced in the lower fluid in a one layer model. The height and radial extent of the hump have been exaggerated. The dotted box indicates the area seen in the photographs.

In order to characterize quantitatively this scattering force, we also investigated interface deformation with an upward beam close to the critical point. An example is illustrated in Figure 10. Associated to a downward deformation due to radiation pressure effects, the interface also deflects upwards, i.e. in the propagation direction, forming a hump whose lateral length scale is much larger than the beam width. Light scattering produces now an upward body force on the liquid within the laser beam, driving an upward flow within the light region. By conservation of fluid mass, this flow is replenished by a downward flow. Since the Reynolds number is small, as illustrated in Figure 4, inertial effects are negligible and purely viscous flows minimize dissipation [77]. The replenishing flow takes the form of a toroidal recirculation whose size is set by the container (see right Inset of Figure 10 for a schematic). Consequently, viscous stresses associated with the recirculation deform the interface upwards and creates a hump whose width is ultimately determined by the size of the cell which is therefore much wider than the laser beam. Assuming a Stokes flow, we can deduce $F_{Scat} \sim \mu \nabla^2 u_0$, where $\mu$ is the viscosity and $u_0$ the induced fluid velocity. Using $\omega_0$ as the optically forced length scale, this leads to $u_0 \sim \chi_T\ P/\mu$, which shows that the flow should be independent of the beam waist. Figure 11a experimentally demonstrates this property. Finally, taking into account buoyancy and the Laplace pressure, as for the description of radiation pressure effects on interfaces, and adding the scattering force, it becomes possible to fit the experimental deformation profiles as shown in Figure 11b.

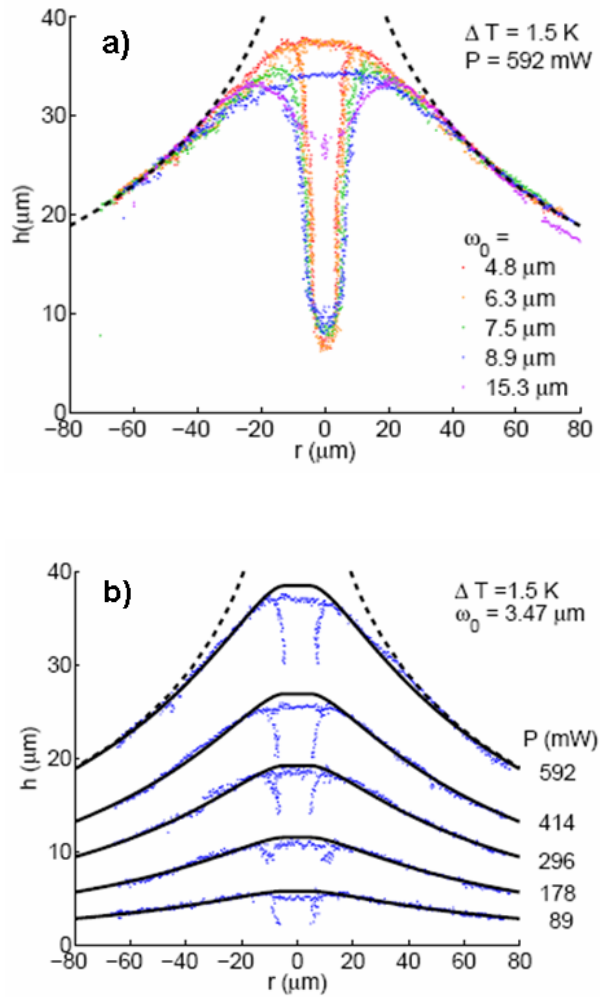

Figure 11 (adapted from Ref. [75]): Interface deformations when the laser shines upwards for $(T - T_C) = 1.5\ K$. The dashed lines represent the pure buoyancy solution, when the Laplace pressure is ignored. a) Variation of the hump profile versus the beam waist $\omega_0$. Away from the beam axis, all the profiles fall onto the same shape, demonstrating that only the laser power, and not $\omega_0$, affects the large-scale hump shape. b) Calculated (solid line) and experimental (dots) hump profile for $(T - T_C) = 1.5\ K$ and $\omega_0 = 4.8\ \mu m$.

## 2.6. Summary

Within a "total-optical-lab-on-a-chip" framework, we demonstrated in Section 2, how the radiation pressure of a laser wave is able to drive large scale flows with flow rates of several hundreds of $\mu m^3/s$ using scattering forces. Such light-induced flows should exist whenever fluids have spatial variation in the refractive index, as in nanocolloidal suspensions [78]. Moreover, we showed that surface effects of radiation pressure are able to create soft channels to guide this flow whose diameter can be tuned by light. This ability to optically actuate liquid flow dynamically in three dimensions, without prefabricated channel, opens a new avenue for building microfluidic devices. Finally, we have presented an electromagnetic instability mechanism of fluid interfaces, still driven by the optical radiation pressure, which leads to optical streaming of fluid interfaces and droplet shedding. The stability of the induced jetting, the actuation of the droplet emission frequency as well as the monodispersity of the produced microdroplets, also offers new perspectives in digital microfluidics. Since these droplets are optically trapped by the exciting beam, they may further be manipulated by light, if required. Consequently, as far as the involved interfacial tensions are weak enough, mechanical effects of light seem very promising to build microfluidic devices or, conversely, to anticipate new optical microfluidic-based micro-systems for fully developed optofluidics applications. In the next Section, we propose a different alternative in order to actuate droplets when interfacial tensions are large enough to prevent efficient actuation by optical forces.

## 3. Laser actuation for digital microfluidics

### 3.1. Mechanical aspects

A far as fast flowing picoliter droplets are used as tight fluid reactors or carriers, new reliable fluid-handling micro-technologies are required to form, convey, synchronize and more generally manipulate accurately these droplets within the robust environment of microchannels. To integrate all these functions onto a single microchip, the method generally consists in plugging different fluid-handling modules, each of theses modules accomplishing one of the required tasks. However, most of these modules need either moving mechanical parts or special micromachining and micropatterning, thus preventing any further spatial reconfiguration in real-time. Moreover, each task needs its solution preventing, up to now, the emergence of a sort of unified approach. Since optical forces are within the $pN$ range, which is often weak to efficiently manipulate fast flowing droplets, we propose an optical alternative based on the production of localized thermocapillary stresses (also known as Marangoni effect [79]) on the droplet interface. Indeed, in presence of surfactant, the magnitude of thermocapillary forces can reach the $\mu N$ range [80], becoming by the way order of magnitudes more efficient than optical forces. Thermocapillary forcing is associated to the temperature dependence of the interfacial tension $\sigma(T)$ between two immiscible fluids. If a droplet flows in a carrying fluid in the presence of a thermal gradient, then this thermal gradient will produce an interfacial tension gradient which will induce in turn a viscous stress in both fluids. It results an interfacial flow, directed toward the area of largest interfacial tension, which develops both inside and outside the drop. Considering mechanical momentum conservation, the droplet will move in the opposite direction. This mechanism is illustrated in Figure 12.

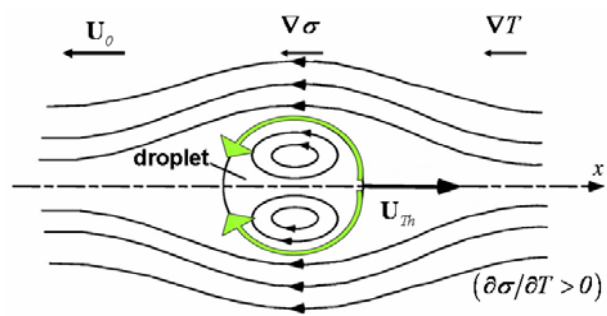

Figure 12: Droplet migration under thermocapillary stresses in a flow field $\vec{U}_o$ when $\partial\sigma/\partial T > 0$. See text for notations.

In an unbounded fluid, the thermocapillary migration of a droplet is given by the following expression [79]:

$$\mathbf{U}_{Th} = -\frac{2}{2\mu_o + 3\mu_i}\left(\frac{\partial\sigma}{\partial T}\right)\frac{R}{2 + \Lambda_i/\Lambda_o}\nabla T, \tag{15}$$

where, $R$ is the droplet radius, $\mu_{o,i}$ and $\Lambda_{o,i}$ are shear viscosities and thermal conductivities, and the subscripts $i, o$ denote the fluids inside and outside the drop. Equation 15 shows that $\mathbf{U}_{Th}$ depends on the thermal gradient and not the temperature directly. Then, even weak, an overheating may drive efficiently thermocapillary flows if the thermal gradient is large, as usually happens with laser light. Moreover, since thermocapillary flows have an interfacial origin, they are particularly suitable to drive flows at small scales where surface effects dominate bulk behaviors, as in microchannels. Such a thermal gradient can be induced by laser, when light is partially absorbed by one of the two fluids. For instance it was shown that a floating drop [30] or a non wetting drop deposited on a substrate [81] can move under laser-driven thermocapillary stresses. Thermocapillary actuation inside microchannels was also demonstrated [82, 83], although the amplitude of the velocity is slightly reduced by wall friction [84]. Finally, note that depending on the sign of $(\partial\sigma/\partial T)$, a laser beam will attract ($\partial\sigma/\partial T < 0$) or repel ($\partial\sigma/\partial T > 0$) a drop. Below, we illustrate the generality of this optical approach by showing how it provides a large set of tools for the contactless manipulation of drops in microchannels when $\partial\sigma/\partial T > 0$ [85].

### 3.2. Experimental setup

The experimental setup, presented in Figure 13, is composed of an inverted microscope (IX 71, Olympus), a continuous $TEM_{00}$ Argon-Ion laser (Innova 300, Coherent) and different types of Polydimethylsiloxane (PDMS) microchannels fabricated by soft lithography techniques. Water (viscosity $\eta_W = 10^{-3}\,Pa.s$, thermal conductivity $\Lambda_W = 0.6\,W/m/K$ at $20\,°C$) and Hexadecane, called hereafter oil, ($\eta_O = 3.34\times10^{-3}\,Pa.s$, $\Lambda_O = 0.14\,W/m/K$) are pumped using syringe pumps; hydrostatic pressure was also used incidentally. A surfactant, Sorbitan monooelate (Span 80), is added to the oil at $2\%\,wt$. Such amount of Span 80, which is quite classical in microfluidics, leads to $(\partial\sigma/\partial T) > 0$ [86], yielding to a thermocapillary force that pushes drops away from the beam. Water droplets are formed in the oil either at T or cross junctions in a central channel; the typical microchannel height is $h = 30 \pm 5\,\mu m$. The flowing water is locally heated with the laser by adding $0.1\%\,wt$ of fluorescein. The corresponding optical absorption of the aqueous phase is $\alpha_a = 1.18\,cm^{-1}$. The beam, injected by the rear port of the microscope and a dichroic mirror (Omega), is focused inside the channel using standard microscope objectives (Olympus) of magnification $\times 2.5$, $\times 5$, $\times 10$ to respectively form beam waist of values $\omega_0 = 10.3, 5.2, 2.6\ \mu m$. In these focusing conditions, the beam can be assumed as cylindrical over the microchannel height. Finally, images of the droplet traffic are acquired with a high speed CMOS camera (Lightning RTD, DRS).

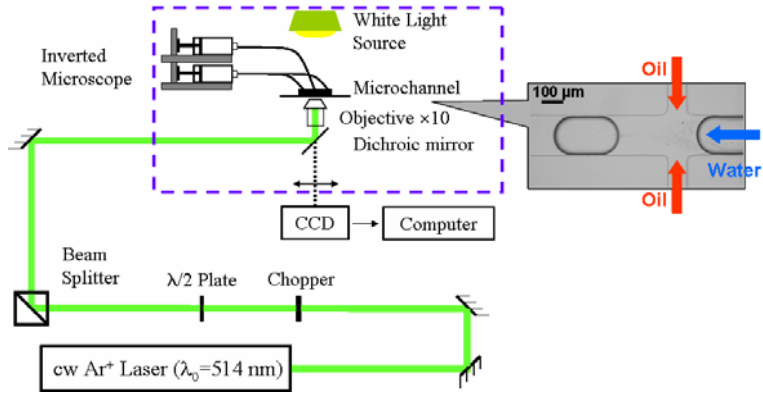

Figure 13: Experimental setup: An Argon-Ion laser is focused through the microscope objective into the microchannel. Syringe pumps control the flow rates of oil and water. Right image: Close view of the water drop formation in a cross microchannel.

### 3.3. Optofluidic mixer: a first step

Fluid mixing in miniaturized systems is a difficult task because diffusion dominates low Reynolds number flows. An appealing method is to use the so-called chaotic advection concept, which is based on the fact that chaotic fluid stream lines can be forced to disperse fluid species effectively, even in smooth and regular flow fields [87]. However, such mixing method often requires the use of complex three-dimensional microstructures. Active mixing, still based on chaotic advection, is usually achieved by periodic perturbation of the flow fields. Several fluid actuation methods were demonstrated using heat convection, differential pressure, ultrasonic actuation, magnetic actuation or electrokinetic pressure; see Ref. [88] for a recent review. One important advantage of these mixers is that they can be activated on-demand. However, compared to passive ones, they need external power sources and hence require more complex packaging. Local heating by laser could bring new insights in active mixing [89] for at least four reasons: (i) Laser waves can easily be focused over spatial scales in the typical range of microchannels; (ii) Laser heating is contactless and thus do not need any complex packaging or patterning of heating elements; (iii) Laser heating is totally reconfigurable in terms of input energy (beam power), spatial extension (beam waist), duration (chopping), and positioning within the channel in a very fast way (compared to hydrodynamic time scales); and finally (iv) multi actuation is also easy to implement using spatial light modulator and/or galvanometric mirrors.

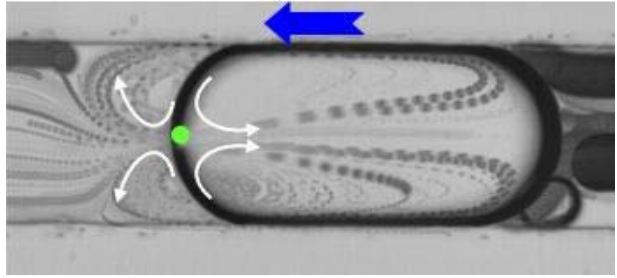

Figure 14 (adapted from Ref. [80]): superposition of 100 frames from a video sequence showing the thermocapillary motion of particle tracers inside and outside a beam-blocked drop. The laser position is indicated by the dot, and the main flow by the wide arrow. The tracer motion is directed toward the hot spot (indicated by the thin arrows), thus demonstrating $(\partial\sigma/\partial T) > 0$. Water flow rate $Q_W = 0.01\ \mu l/mn$, oil flow rate $Q_O = 0.1\ \mu l/mn$, frame rate $200\ Hz$, beam power $P = 60\ mW$, beam waist $\omega_0 = 5.2\ \mu m$, microchannel section $140 \times 35\ \mu m^2$, carbon tracers: $\varnothing \approx 1\ \mu m$.

A first step toward thermocapillary mixing in microchannel is illustrated in Figure 14. A water drop flowing in a $140\ \mu m$ wide channel at a few $mm/s$ is blocked by the laser due to thermocapillary stresses; the blocking mechanism is discussed in the following subsection. Using tracer particles in both fluids, we observe convection rolls inside and outside the drop due to the presence of the laser beam. In the present experiment the tracer velocity at the beam location is $11\ mm/s$. The presence of rolls inside and outside the drop illustrates the procedure to perform thermocapillary mixing in both fluids. Indeed, a single set of vortices is obviously not sufficient, but chaotic mixing should emerge by using three independent beams, each producing its set of vortices.

### 3.4. Optofluidic valve

Experiments were performed in a cross microchannel. In the example illustrated in Figure 15, water droplets are formed by hydrodynamic focusing and are emitted at a frequency of $0.5\ Hz$. Their mean length and velocity are respectively $L = 330\ \mu m$ and $U_0 = 1.2\ mm/s$. When the front of the forming drop reaches the beam spot position it is pushed away from the laser by thermocapillary forces; see Equation 15 using $(\partial\sigma/\partial T) > 0$. Since it is constrained by the channel walls, the interface cannot turn around the beam which thus behaves as a sort of "soft" wall. When the beam power reaches a threshold (such as $|U_{Th}| \geq U_0$), the drop is blocked in the microchannel.

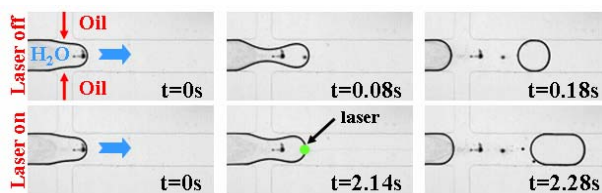

Figure 15 (adapted from Ref. [80]): Microfluidic valve: In a cross-shaped microchannel, the oil and water flows respectively enter from the lateral and the central channels. In the absence of laser excitation, drops are shed with a typical break-off time of *100 ms* . When the laser is applied, the interface is blocked for several seconds, producing a larger drop. Water flow rate $Q_W = 0.08$ *μl/mn* , oil flow rate $Q_O = 0.9$ *μl/mn* , beam power $P = 80$ *mW* , beam waist $\omega_0 = 5.2$ *μm* , and main channel width *200 μm* and lateral channels width *100 μm* .

When the microchannel is pumped at constant flow rates, the blocking time $T_b$ increases with the incident beam power [80]. This is illustrated in Figure 16 for different blocking positions. $T_b$ also decreases with the beam waist at a given beam power and significantly increases when flows are controlled by pressure instead of flow rate [90].

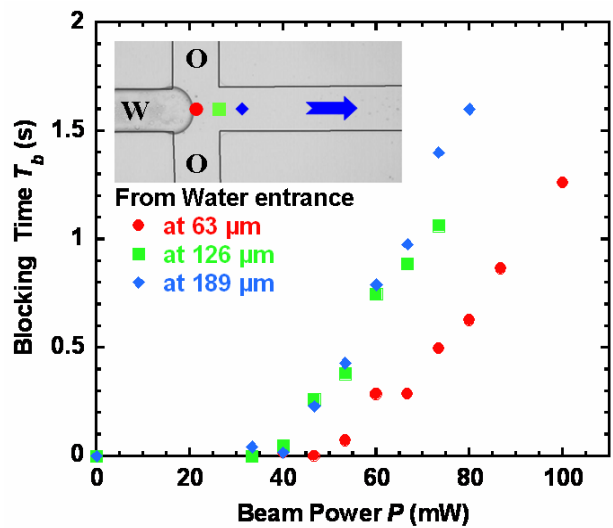

Figure 16 (adapted from Ref. [80]): Dependence of the blocking time $T_b$ on laser power and position indicated in the picture. Water flow rate $Q_W = 0.03$ *μl/mn* , oil flow rate $Q_O = 0.1$ *μl/mn* , beam waist $\omega_0 = 2.6$ *μm* , and channel widths *125 μm* .

*3.5. Optofluidic sampler*

Another way to calibrate droplets consists in using as a drain, a T junction with two arms of different length [91]. At the T junction, the flowing drop divides into two daughter drops whose lengths are imposed by the asymmetry of the two exit channels. While very efficient and reliable, this type of sampler is passive and the asymmetry in daughter droplets can only be varied by switching to channels with different asymmetry. This type of asymmetry can be optically actuated at a classical T junction by positioning the laser close to one of the outlets. This is illustrated in Figure 17.

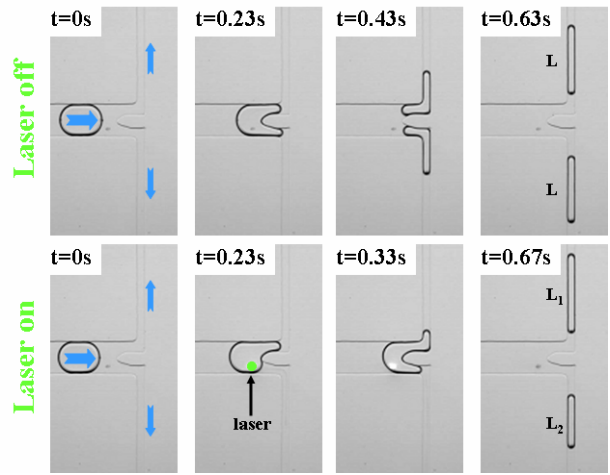

Figure 17 (adapted from Ref. [85]): Microfluidic sampler. In the absence of laser the mother drop divides into two daughters of same size. Laser forcing allows for breaking this symmetry in a controlled fashion. Water flow rate $Q_W = 0.02\ \mu l/mn$, oil flow rate $Q_O = 0.2\ \mu l/mn$, beam power $P = 87\ mW$, beam waist $\omega_0 = 5.2\ \mu m$, and main channel width $200\ \mu m$.

By momentarily blocking the drop flow close to one outlet, the drop preferentially flows in the other exit. We then obtained daughter droplets of different sizes. This asymmetry, discussed in Ref. [85], is characterized from the mean aspect ratio $\Lambda = (L_1 - L_2)/(L_1 + L_2)$, where $L_1$ and $L_2$ are the lengths of the two daughter droplets. The extreme values $\Lambda = 0,\ 1$ respectively correspond to the symmetric case and a situation where the mother drop is entirely diverted toward a single outlet. Both situations are illustrated in Figure 18. When the beam power is smaller than a threshold, still defined by $|U_{Th}| \geq U_o$, the mother drop divides in two daughters whose aspect ratio increases with $P$. Above threshold, equal to $P = 100\ mW$ for the experiment presented in Figure 17, the mother does not divide anymore and flows toward the unblocked outlet. Below the beam power threshold the device behaves as a sampler while above it becomes a droplet switcher.

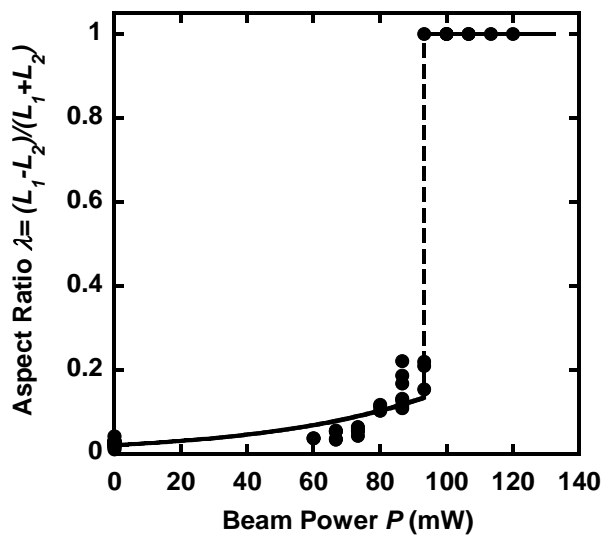

Figure 18 (adapted from Ref. [85]): Microfluidic sampler and switcher: dependence of the asymmetry in daughter droplet sizes on laser power ( $\Lambda = 0$ is for symmetric drops and $\Lambda = 1$ is for drop switching). Lines are guides for the eye. The operating conditions are those of Figure 17.

### 3.6. Optofluidic switcher

A second approach to microfluidic switching can be devised using slightly different channel geometry. Instead of blocking droplets at the exit, we can just switch their trajectories at an enlargement of the main channel. This is illustrated in Figure 19 using a Y-type channel. The main advantage of this approach is that droplet can be manipulated at velocities as large as the $cm/s$ [92] because blocking is not required anymore. On the one hand, without laser, water drops formed upstream arrive in the enlarged part of the Y junction and flow asymmetrically downstream toward the exit channel of smaller hydrodynamic resistance (the down one in pictures). This weak asymmetry is forced by shortening one of the outlets, as in Ref. [91]. On the other hand, when a flowing drop is shined below its equator by the laser, it is suddenly pushed out toward the opposite direction above a beam power threshold. Once again, the beam behaves as an active wall whose stiffness is optically actuated.

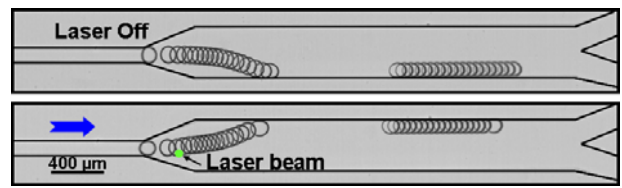

Figure 19 (adapted from Ref. [92]): Microfluidic switcher: Superposition of successive frames illustrating drop switching by local thermocapillary actuation. The dot indicates the laser location observed by the fluorescence of the water-dye solution. Water flow rate $Q_W = 0.05 \ \mu l/mn$ , oil flow rate $Q_O = 5 \ \mu l/mn$ , $U_0 = 2.6 \ mm/s$ , $R = 60 \ \mu m$ , $P = 58 \ mW$ , frame rate $100 \ fps$ . Channel width: inlet $100 \ \mu m$ , enlargement $400 \ \mu m$ and exits $200 \ \mu m$ .

The beam power variation of the droplet switching efficiency was investigated in Ref. [92]. It presents a very steep behavior. Below a power threshold, all drops exit from the asymmetrically forced outlet while above threshold they all deviate toward the opposite outlet, as illustrated by the beam power variation of the drop trajectories in the microchannel (Figure 20).

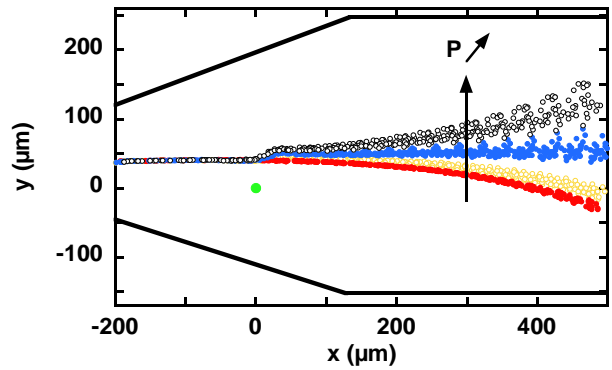

Figure 20 (adapted from Ref. [92]): Trajectory of the drop center of mass for various beam powers (from bottom to top: $P = 0, 48, 53, 58\ mW$); switching occurs at $P = 53\ mW$. The beam location is indicated by the dot at $(x = 0, y = 0)$. Water flow rate $Q_W = 0.05\ \mu l/mn$, oil flow rate $Q_O = 5\ \mu l/mn$, $U_0 = 2.6\ mm/s$, $R = 60\ \mu m$.

Due to the expression of the thermocapillary velocity (Equation 15), the beam power threshold varies with the drop velocity and radius. Generally speaking, it appears that switching occurs when the mean deviation angle between the natural and the beam modified trajectory reaches some threshold value associated to the forced asymmetry in the $y$ direction for the droplet traffic in the absence of laser. In the above experiment, the mean deviation angle over the interaction time is $3°$; this mean critical angle and the related beam power required for switching could be even lowered by locating the beam closer to the channel center line.

### 3.7. Optofluidic sorter

To demonstrate the reliability of the above laser switching process for sorting droplets on chips, we used the same Y-type microchannel with two upstream T junctions (seen on the left of Figure 21) instead of one. The very left one is used for the generation of pure water droplets and the other produces water-dye droplets as in the above switching experiment. An asymmetry in outlets, designed as in Ref. [91], still forces the drops to naturally flow toward the lower outlet, whatever their composition. Since the residual optical absorption of water at $\lambda_0 = 514.5\ nm$ is negligible ($\approx 3 \cdot 10^{-3}\ cm^{-1}$), pure water droplets are rather insensitive to the beam while water-dye droplets are spontaneously deviated and exit from the other outlet at beam powers above threshold, as illustrated in Figure 21.

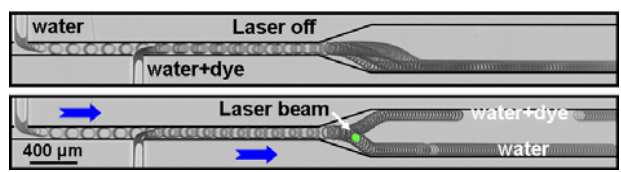

Figure 21 (adapted from Ref. [92]): Microfluidic sorting: Top picture: frames superposition showing three drops (water-dye + water + water-dye) which are naturally drawn toward the lower exit channel without laser. Bottom picture: frames superposition showing sorting of a pure water and a water-dye drop when the laser shines droplets. The dot indicates the laser location observed by the fluorescence of the water-dye solution. Water flow rate $Q_W = (0.03 + 0.03)$ $\mu l/mn$, oil flow rate $Q_O = 5$ $\mu l/mn$, $U_0 = 2.2$ $mm/s$, $R = 59\mu m$ and $P = 159$ $mW$, frame rate $100$ $fps$. Channel width: inlets $100$ $\mu m$, enlargement $400$ $\mu m$ and exits (not shown) $200$ $\mu m$.

### 3.8. Optofluidic merger

Finally, beyond mixing inside and outside flowing drops, there is a second type of mixing process required for digital microfluidics. As far as flowing droplets are used as micro-reactors, the coalescence of at least two daughter droplets, supporting different reagents, is necessary. This coalescence stage is difficult to accomplish passively in microchannels because surfactant molecules are known to stabilize drops against merging, even if a recent work has shown a very interesting alternative to circumvent this difficulty [93]. Active control can be efficient too, as illustrated by recent investigations on charged microdroplets [94].

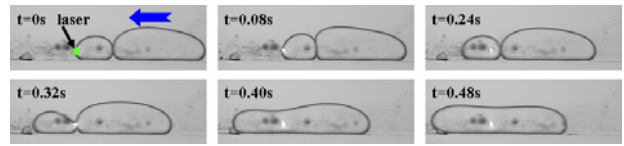

Figure 22 (adapted from Ref. [85]): Time sequence showing droplet fusion by laser-induced thermocapillary stresses. Half of a $200$ $\mu m$ wide channel is presented, droplets flowing close to one edge. Coalescence occurs when the beam intercepts the touching interfaces. Water flow rate $Q_W = 0.2$ $\mu l/mn$, oil flow rate $Q_O = 0.9$ $\mu l/mn$, beam power $P = 67$ $mW$, beam waist $\omega_0 = 2.6$ $\mu m$.

The thermocapillary coupling may also circumvent the lubrication film problem by evacuating the surfactant molecules, as illustrated in Figure 22. Here, the downstream drop is held stationary by the thermocapillary blocking until a second one collides with it (Figure 22, t=0s). At this point, the two drops advance until the laser intercepts the adjacent interfaces, and we observe that the lubrication film between the drops is rapidly evacuated, leading to merging [85].

### 3.9. Summary

Within a "total-optical-lab-on-a-chip" framework, we proposed in Section 3 a method, thermocapillary actuation by laser, to optically manipulate droplets inside a microchannel. The method seems promising because it takes advantage of the dominance of surface effects in microfluidics, as opposed to dipolar couplings used by optical trapping standards. Within this framework, we developed a basic optical toolbox for digital microfluidics, including valves, samplers, switchers, sorters, mergers and a first step towards mixers. These drop-handling functions can also be used in combination; a first example is illustrated in Ref. [85]. As any optical technique, our method is still contactless, allowing droplet manipulation without moving parts or dedicated microfabrication. Optical forcing also allows reconfigurability in real time, either spatially or temporally, with one or many beams. Finally, extension to near IR excitation would make the method even more flexible and amenable for applications since the optical absorption of pure water at $\lambda_o = 1500\ nm$ is $\alpha_a \approx 20\ cm^{-1}$, thus eliminating polluting dyes for heating and reducing the required beam power range by a factor twenty.

## 4. Conclusion and prospects

The main purpose of this survey was to theoretically and experimentally explore fluid actuation by light within a "total-optical-lab-on-a-chip" framework. To implement such a strategy in microchannels, we need interface forcing mechanisms. In a first part (Section 2), we demonstrated how the optical radiation pressure of a laser wave was able to induce large scale flows, to create soft channels for guiding these flows, and to drive optical streaming of fluid interfaces and droplet shedding. In Section 3, we implemented an indirect optical interface coupling, opto-capillarity, to build an optical toolbox for digital microfluidics. The method shows appealing prospects to form valves, samplers, switchers, sorters, mergers, and mixers, and to even develop further these actuators and to couple them. Drops were nevertheless formed by flow focusing instead of light.

These experimental results bring us back to the Glückstad concept of "optical chip" [25], where flows, channels and operating functions on microdroplets would all be performed optically on the same device, extending by the way the optofluidic approach [16, 17] to digital microfluidics.

On the one hand, the mechanisms at the origin of flows driven by bulk radiation pressure effects, using the scattering force, are analogous to those involved in the recent development of the so-called optical chromatography of microparticles, which is a relatively new technique of laser separation [95] still under development [96, 97]. Since in our experiment in near-critical fluids the particle-like fluctuations are only $100\ nm$ large with a very weak refraction index contrast, and since refractive index contrast in real colloids is usually much larger, we expect an enhancement of scattering forces in classical suspensions of nanoparticles. Thus, bulk radiation forces should be able to push collectively a suspension as a whole and this pushing effect should be still very efficient at nanometric scales. While previous calculations performed in the geometric optics approximation (which is not very accurate for these sizes) gave $100\ nm$ as the smallest size for separation by optical chromatography [95], using instead the Rayleigh cross-section, we find $F_{Scat}=0.5\ pN$ for a glass bead of radius $100\ nm$ which may lead to a water flow of mean velocity $0.2\ mm/s$. We can therefore conclude that the scattering force should initiate a flow in a suspension of colloidal particles of a few tens of $nm$ and would become particularly efficient in the $100\ nm$ range to initiate and sustain microfluidics flows.

The interfacial component of the radiation pressure efficiency would nevertheless suffer from a drawback: interfacial tension must be weak enough to allow significant mechanical effects of light. Hopefully, digital microfluidics frequently uses surfactants at concentration generally above the critical micellar concentration (CMC). Is it sufficient to stabilize liquid channels? Since the resulting interfacial tensions are within the range of *0.1 − 1 mN/m* , radiation pressure may stabilize small radius channels as far as these radii are reachable (remember the subcritical behaviour of the channel radius versus the incident beam power, Figures 8-9). Conversely, is it sufficient to destabilize fluid interfaces? Up to now, the answer is probably no due to the fact that even more power will be required to induce a liquid jet. A significant power increase also means an increase of the unavoidable side effects such as those triggered by laser heating. On the other hand, the major limitation of the opto-capillary approach is its dissipative nature (use of thermal gradients triggered by optical absorption). Its application viability would need an extension towards near IR in order to (i) eliminate polluting dyes and, (ii) significantly reduce the required beam powers in order to directly integrate the laser on the chip. These requirements can easily be achieved. However, we will not be able to use the same laser (multiplexed as much as needed) to control the flow and simultaneously actuate the droplet shedding because the required beam power ranges are incompatible to each other at the same wavelength. Typically, too much IR light, to trigger surface effects of radiation pressure, will vaporize fluids while visible light will be inefficient for actuating droplets. Consequently up to now the best compromise, if it exists, would probably be the use of different lasers, each accomplishing his task. The subject clearly deserves further investigations!

As a preliminary conclusion, the present survey illustrates another step in the exploration of the coupling between optics and microfluidics. Associated to much more exhaustive reviews [16, 17, 18] and to the increasing number of publications in this domain, it illustrates the emergence of a new and very active research field. In fact, the development of optofluidics, and in our case optohydrodynamics, is obviously in its infancy, probably because it intimately mixes disciplines which have been considered far away for a long time. Right now, everything is open since we even don't feel the limits of our framework. The constant progress in microfabrication always pushes these limits, engaging more and more fundamental and applied investigations. Our presentation participates to this dash, giving one more view of the synergy between optics and fluidics and allowing us to dream on "optical chips".

**Acknowledgments**

The authors gratefully acknowledge the Conseil Régional d'Aquitaine Contract No 2004025003 (J.P.D., D.L., H.C., R.W.) and Contract No. 20061102030 (J.P.D., M.R.S.V., R.W.), the Centre National de la Recherche Scientifique PICS No 3370-2005 (J.P.D.), the National Science Foundation (R.D.S.), NSF Contract No. CBET-0730629 (W.W.Z.) for financial support, and the Rhodia Laboratory of Future for access to microfabrication facilities. J.P.D. wish to thank Iver Brevik, Charles Baroud and Jean-Marc Fournier for helpful discussions, Joël Plantard for technical assistance, and Jean-Baptiste Salmon for advice and technical assistance in microfabrication.